\documentclass[onecolumn,tightenlines,nofootinbib,preprintnumbers,amsmath,amssymb]{revtex4}
\usepackage[usenames,dvipsnames]{color}
\usepackage{caption2}
\usepackage{dcolumn}
\usepackage{makecell}
\usepackage{graphicx}% Include figure files
\usepackage{subfigure}
\usepackage{epstopdf}
\usepackage{bm}% bold math
\usepackage{lscape}
\usepackage{slashed}
\usepackage{booktabs}
\renewcommand{\arraystretch}{0.9}
\usepackage{setspace}

\begin{document}
\begin{spacing}{1.5}
%\nofiles
\title{CP asymmetry from resonance effect of B meson decay process with $\pi$ and K final states}
% Force line breaks with \\
\author{Gang L\"{u}$^{1}$\footnote{Email: ganglv66@sina.com}, Xi-Liang Yuan$^{1}$\footnote{Email: m18937315007@163.com}, Na-Wang$^{1}$\footnote{Email: wangna@haut.edu.can}, Xin-Heng Guo $^{2}$\footnote{Email: xhguo@bnu.edu.cn}}
\affiliation{\small $^{1}$ Institute of Theoretical Physics, College of Science, Henan University of Technology, Zhengzhou 450001, China\\
\small $^{2}$ College of Nuclear Science and Technology, Beijing Normal University, Beijing 100875, China\\}
%\date{\today}
\begin{abstract}
We introduce the new resonance of $V\rightarrow K^{+}K^{-}$ $(V=\phi, \rho, \omega)$, which produces some new strong phase associated with vector meson resonance and thus can cause relatively large CP asymmetry at the range of interferences. There are the resonances of $\phi \rightarrow K^{+}K^{-}$, $\rho \rightarrow K^{+}K^{-}$ and $\omega \rightarrow K^{+}K^{-}$ due to the mixing of vector mesons $\phi$, $\rho$, $\omega$. We calculate the CP asymmetry from the decay modes of $B \rightarrow KK\pi(K)$. Meanwhile, the localised CP asymmetries are presented and some detailed analysis can be found. The CP asymmetry from the decay mode of ${B}^{-}\rightarrow \phi\pi^{-}\rightarrow K^{+}K^{-}\pi^{-}$ is also presented in our framework which is well consisted with LHC experiment. The introduced CP asymmetry can provide a favorable theoretical support for the experimental exploration in the future.
\end{abstract}
\maketitle

\section{\label{intro}Introduction}
CP asymmetry has been researched for many years in the field of particle physics which firstly originated in 1964 and has attracted wide attention \cite{0JJ1964}. C and P refer to charge conjugation transformation and parity inversion transformation, respectively \cite{TGe2017}. In order to find new physical signals and search the standard model (SM), CP asymmetry is extremely important platform.
Now, various of studies about the B meson from the theory and experiment play more and more important role and have been increasing maturity, which can search the standard model and reveal the quality of the interaction between mesons.
Compared to other mesons, the B meson contains the heavy b quark and the decay process of it is very suitable for detecting the CP asymmetry since the perturbation result is great.
In this work, we consider the new resonance of $V\rightarrow K^{+}K^{-}$ $(V=\phi, \rho, \omega)$, which produces new strong phase associated with vector meson resonance and thus can cause relatively large CP asymmetry at the range of interferences. The vector meson resonances not only provide rich physical mechanism for particle property research but also show rich informations of intermediate meson in multi-body decay process \cite{MDA1979}.

Theoretically, in order to investigate the popular non-leptonic weak decay of B meson, methods such as perturbative QCD (PQCD) \cite{P2017} and QCD factorization (QCDF) \cite{N2003} have been fully explored and widely used by researchers.
The perturbation QCD method is used to separate out the hard portion due to transverse momentum where QCD correction owe participated and dealed with it by perturbative theory. The Sudakov factor is introduced to depress the non-perturbative effects. The contribution of non-perturbation is involved in hadron wave function. Under the heavy quark limit from the framework of QCDF, if the mass of the quark is very heavy (such as b quark) in the weak decay process, we can make the mass tend to infinity so as to ignore the contribution of high order 1$/$$m_b$. The two-body non-leptonic decay amplitude can be expressed  as the product of the form factor from the initial state meson to the final state meson and the light cone distribution amplitude of the other final state meson in the heavy quark limit. The non-leptonic three-body decay problem of the B meson can be solved, which has become one of the methods widely used by researchers in recent years \cite{Qi2019}. It is a good way to study CP asymmetry from resonance effect for B meson decay process with $\pi$ and K final states under QCDF \cite{1cab,2lu2021}.

Vector meson dominant model (VMD) forecasts that the vacuum polarisation of the photon is made up of vector mesons of $\rho^{0}(770)$, $\omega(782)$ and $\phi(1020)$ \cite{NB1967}. The two-pion state dominates the photon couples to the neutral vector meson when $e^{+}e^{-}$ decay into the pair of $\pi^{+}\pi^{-}$. The transitions of $\omega(782)$($\phi(1020)$) and $\rho^{0}(770)$ decay to $\pi ^+\pi ^-$ pair which originate in isospin breaking and isospin conservation related to the mixings of $\rho^{0}(770)-\omega(782)$ and $\rho^{0}(770)-\phi(1020)$ \cite{2lu2022}. Through the unitary matrix, the physical state of the intermediate state and the isospin state can be combined. The dynamics mechanism can be obtained from the interference of $\phi(1020)$, $\rho^{0}(770)$ and $\omega(782)$ mesons. The new strong phase is formed, which may have an effect on the CP asymmetry of hadron decay under the framework of the intermediate resonance hadrons.

As a whole, we make a brief introduction including the research background and situation in this part. In second part, we analyse the resonance effect in the three-body decay process from the physical mechanism and the detailed formalisms in section A. Then we make a comprehensive analysis of CP asymmetry in the decay processes of $\bar{B}^{0}\rightarrow K^{+}K^{-} \bar K^{0}(\pi^{0})$ and $B^{-}\rightarrow K^{+}K^{-}K^{-}(\pi^{-})$ under the vecor mesons resonance mechanism of $\phi \rightarrow K^{+}K^{-}$, $\rho\rightarrow K^{+}K^{-}$ and $\omega \rightarrow K^{+}K^{-}$ in section B and section C, which is the main part of this work.
Then we take the decay process of $\bar B^{0}\rightarrow\phi\bar K^{0}\rightarrow K^{+}K^{-}\bar K^{0}$ as an example to illustrate the form of the decay amplitude after considering above resonance effect and present the typical three-body decay amplitude in the third part. The numerical results about the analysis of CP asymmetry and the localised form are given in the fourth part. We make a summary and conclusion in the last part.

\section{\label{sum}THE ANALYSIS OF THREE-BODY DECAY PROCESS}
\subsection{\label{subsec:form}The mixing mechanism}
Based on the model dominated by vector mesons, the vector mesons of $\phi(1020)$, $\rho^{0}(770)$ and $\omega(782)$ are polarized by the photon which is generated from the annihilation of the positive electron and negative electron. They can decay into $K^{+}K^{-}$ pairs. Based on the isospin fields $\phi_{I}$, $\rho^{0}_{I}$, $\omega_{I}$, we can construct the physical particle states. The two representations are related to each other by the unitary matrix R \cite{2lu2022}.

Since the resonance effect is not from the physical field, thus the unitary matrix R is required to change the isospin field into the physical field. The relationship between the isospin fields ($\phi_{I}$, $\rho_{I}^{0}$, $\omega_{I}$) and the physical fields ($\phi$, $\rho^{0}$, $\omega$ ) is connected by the the unitary matrix R which can be written as follows \cite{6Lu2017}:
\begin{equation}
R  =
\left (
\begin{array}{lll}
<\rho_{I}|\rho> & \hspace{0.5cm} <\omega_{I}|\rho>  &\hspace{0.5cm}<\phi_{I}|\rho>\\[0.5cm]
<\rho_{I}|\omega> &  \hspace{0.5cm}<\omega_{I}|\omega>&\hspace{0.5cm}<\phi_{I}|\omega>\\[0.5cm]
<\rho_{I}|\phi>&\hspace{0.5cm} <\omega_{I}|\phi> & \hspace{0.5cm} <\phi_{I}|\phi>
\end{array}
\right )\\
=
\left (
\begin{array}{lll}
 ~~~~1 &\hspace{0.5cm} -F_{\rho\omega}(s) &
\hspace{0.5cm}  -F_{\rho\phi}(s)\\[0.5cm]
\displaystyle  F_{\rho\omega}(s) &  \hspace{0.5cm}~~~ 1 &
\hspace{0.5cm} \displaystyle - F_{\omega\phi}(s)\\[0.5cm]
\displaystyle   F_{\rho\phi}(s)
& \hspace{0.5cm}  \displaystyle   F_{\omega\phi}(s)& \hspace{1.0cm} 1
\end{array}
\right ),
\label{L2}
\end{equation}
where $F_{\rho\omega}(s)$, $F_{\rho\phi}(s)$, $F_{\omega\phi}(s)$ is order $\mathcal{O}(\lambda)$, $(\lambda\ll 1)$ \cite{2lu2022}. The physical state of this transformation can be expressed as:
$\rho^{0}=\rho^{0}_{I}-F_{\rho\omega}(s)\omega_{I}-F_{\rho\phi}(s)\Phi_{I}$,
$\omega=F_{\rho\omega}(s)\rho^{0}_{I}+\omega_{I}-F_{\omega\phi}(s)\Phi_{I}$,
$\Phi=F_{\rho\Phi}(s)\rho^{0}_{I}+F_{\omega\phi}(s)\omega_{I}+\Phi_{I}$.

In view of representations from the physics and isospin, we make the definitions of propagator as $D_{V_1V_2}=\left< 0|TV_1V_2|0 \right> $ and $D_{V_1V_2}^{I}=\left< 0|TV_{1}^{I}V_{2}^{I}|0 \right>$. $V_{1}$ and $V_{2}$ of $D_{V_{1}V_{2}}$ refer to the meson of $\rho^{0}$, $\omega$ and $\phi$, respectively. In fact, $D_{V_{1}V_{2}}$ is equal to zero since there is no three vector meson mixing under the physical representation. Besides, according to the expression under the physical state of the three vector meson mixing, the parameters of $F_{\rho\omega}$, $F_{\rho\phi}$ and $F_{\omega\phi}$
are order of $\mathcal{O}(\lambda)$ ($\lambda\ll 1$). There is higher order terms from the contributions of above two terms or more terms which can be safely neglected \cite{2lu2022}. As a consequence, we can get \cite{15Li2022}:
\begin{eqnarray}
F_{\rho\omega}=\frac{\Pi_{\rho\omega}}{S_{\rho}-S_{\omega}},~~F_{\rho\phi}=\frac{\Pi_{\rho\phi}}{S_{\rho}-S_{\phi}},
~~F_{\omega\phi}=\frac{\Pi_{\omega\phi}}{S_{\omega}-S_{\phi}},
\end{eqnarray}
where $\Pi_{\rho\omega}$, $\Pi_{\rho\phi}$ and $\Pi_{\omega\phi}$ are the mixing parameters. Then we define
\begin{eqnarray}
\widetilde{\Pi}_{\rho\omega}=\frac{S_{\rho}\Pi_{\rho\omega}}{S_{\rho}-S_{\omega}},
~~\widetilde{\Pi}_{\rho\phi}=\frac{S_{\rho}\Pi_{\rho\phi}}{S_{\rho}-S_{\phi}},
~~\widetilde{\Pi}_{\omega\phi}=\frac{S_{\omega}\Pi_{\rho\omega}}{S_{\omega}-S_{\phi}},
\end{eqnarray}
where $S_{V}$ and $m_{V}$ refer to the inverse propagator and  the mass of vector meson $V$ $(V=\phi, \rho, \omega)$, respectively. The propagator $S_{V}$ is associated with the invariant mass $\sqrt s$ which can clearly reflect the value of CP asymmetry.
From above equations, we obtain $\tilde{\Pi}_{\rho \omega}$, $\tilde{\Pi}_{\rho \phi}$ and $\tilde{\Pi}_{\omega \phi}$ after definition about $\tilde{\Pi}_{V_1V_2}=\frac{S_{V_1}\Pi _{V_1V_2}}{S_{V_1}-S_{V_2}}$. Such as we use $\frac{S_{\rho}\Pi_{\rho\omega}}{S_{\rho}-S_{\omega}}$($\frac{S_{\omega}\Pi_{\omega\rho}}{S_{\omega}-S_{\rho}}$) to define $\widetilde{\Pi}_{\rho\omega}$($\widetilde{\Pi}_{\omega\rho}$). The same is true for the conversion of the mixing parameters of $\widetilde{\Pi}_{\omega\phi(\phi\omega)}$ and $\widetilde{\Pi}_{\rho\phi(\phi\rho)}$.
The mixing parameters of $\widetilde{\Pi}_{\rho\omega}(s)$, $\widetilde{\Pi}_{\rho\phi}(s)$ and $\widetilde{\Pi}_{\omega\phi}(s)$
is the momentum dependence for $\rho-\omega$, $\rho-\phi$ and $\omega-\phi$ interferences \cite{8Lu2011,MN2000}.
The precise value of $\rho$, $\omega$ and $\phi$ mixing parameters are given by considering the parameters which have been measured by Wolfe and Maltman \cite{P2009,P2011}. On the basis of $\widetilde{\Pi}_{\rho\omega}(s)={\mathfrak{Re}}\widetilde{\Pi}_{\rho\omega}(m_{\omega}^2)+{\mathfrak{Im}}\widetilde{\Pi}_{\rho\omega}(m_{\omega}^2)$, $\widetilde{\Pi}_{\rho\phi}(s)={\mathfrak{Re}}\widetilde{\Pi}_{\rho\phi}(m_{\phi}^2)+{\mathfrak{Im}}\widetilde{\Pi}_{\rho\phi}(m_{\phi}^2)$ and $\widetilde{\Pi}_{\omega\phi}(s)={\mathfrak{Re}}\widetilde{\Pi}_{\omega\phi}(m_{\phi}^2)+{\mathfrak{Im}}\widetilde{\Pi}_{\omega\phi}(m_{\phi}^2)$, then we can present: $\mathfrak{Re}\widetilde{\Pi}_{\rho\omega}(m_{\omega}^2)=(-4760\pm440)
\rm{MeV}^2$, ${\mathfrak{Im}}\widetilde{\Pi}_{\rho\omega}(m_{\omega}^2)=(-6180\pm3300)
\textrm{MeV}^2$; $\mathfrak{Re}\widetilde{\Pi}_{\rho\phi}(m_{\phi}^2)=(796\pm312)
\rm{MeV}^2$, ${\mathfrak{Im}}\widetilde{\Pi}_{\rho\phi}(m_{\phi}^2)=(-101\pm67)
\textrm{MeV}^2$; $\mathfrak{Re}\widetilde{\Pi}_{\omega\phi}(m_{\phi}^2)=19000
\rm{MeV}^2$, ${\mathfrak{Im}}\widetilde{\Pi}_{\omega\phi}(m_{\phi}^2)=(2500\pm300)
\textrm{MeV}^2$.

\subsection{\label{subsec:form} The formalism of CP asymmetry}
We take the $\bar B^{0}\rightarrow \phi(\omega,\rho^0) \bar K^{0}\rightarrow K^{+}K^{-} \bar K^{0}$ decay channel as an example to study CP asymmetry. In the diagram (a) of Fig.1, $\bar B^{0}$ meson decays into $ \bar K^{0}$ and $K^{+}K^{-}$ pair which is produced directly by $\phi$ resonance effect. Meanwhile, it is known that $K^{+}K^{-}$ pair can also exist from the intermediate state of $\omega$ or $\rho$ meson. Therefore, we also consider the processes of $\rho$ and $\omega$ decay into $K^{+}K^{-}$, which are shown in the diagrams of (d) and (g) of Fig.1. Compared with above decay processes, the diagram (b) is different since $\phi$ meson decays into $K^{+}K^{-}$ through $\rho$ resonance and the mixing parameter $\rho-\phi$ is given here. The diagrams of (c), (e), (f), (h) and (i) are similar to the diagram (b). However we find there are the resonance effects of $\rho-\phi$($\phi-\rho$), $\omega-\phi$($\phi-\omega$) and $\rho-\omega$($\omega-\rho$) which produce $\Pi_{\rho\phi(\phi\rho)}$, $\Pi_{\omega\phi(\phi\omega)}$, $\Pi_{\rho\omega(\omega\rho)}$ mixing parameters, respectively.

\begin{figure}[h]
\centering
\includegraphics[height=8cm,width=12cm]{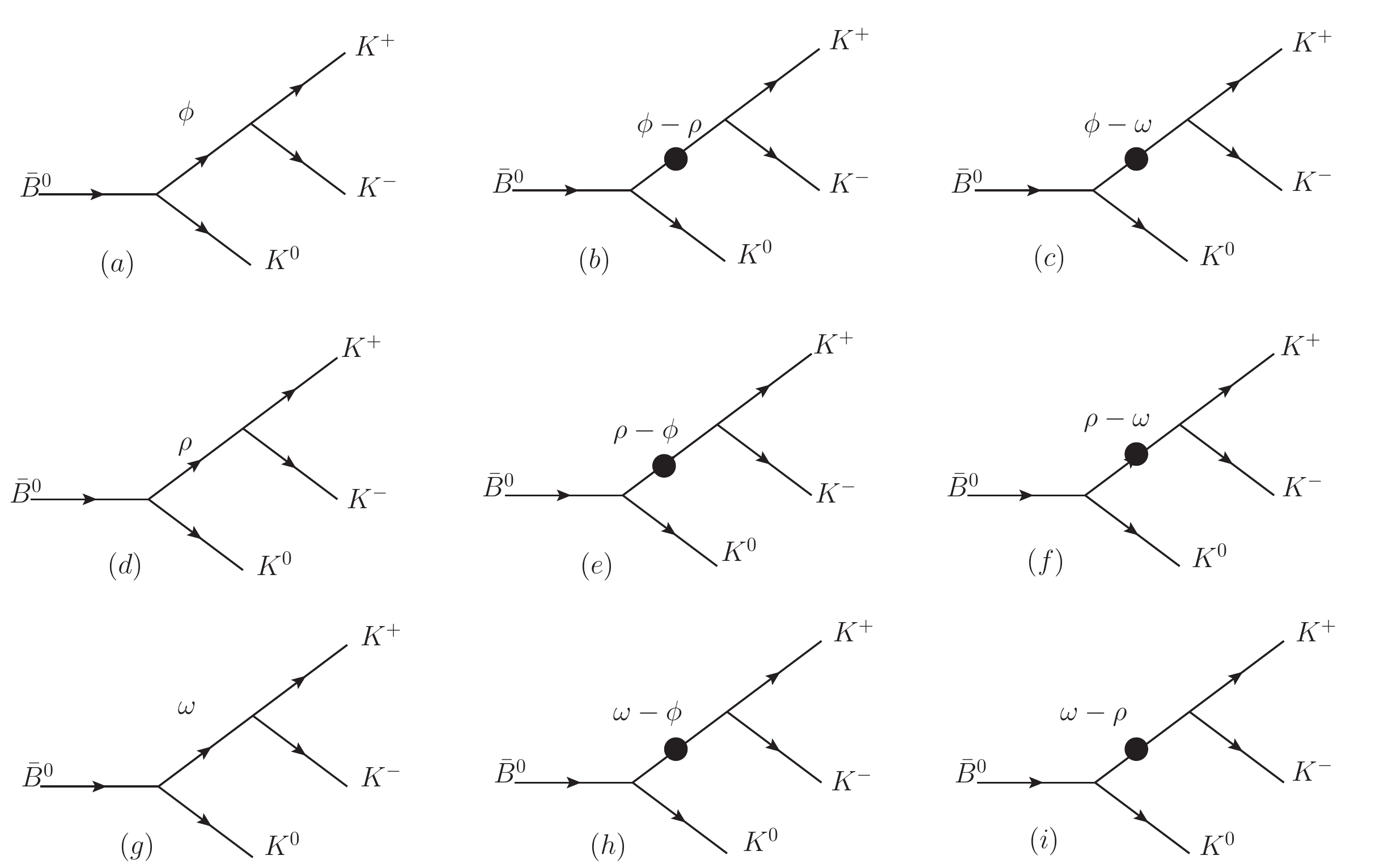}
\caption{The decay diagrams for the channel of $\bar B^{0}\rightarrow K^{+}K^{-} \bar K^{0}$.}
\label{fig1}
\end{figure}

The amplitude A of the three-body decay process of $\bar B^{0}\rightarrow K^{+}K^{-} \bar K^{0}$ can be expressed as:
\begin{eqnarray}
A=\big<K^{+}K^{-} \bar K^{0}|H^T|\bar B^{0}\big>+\big<K^{+}K^{-} \bar K^{0}|H^P|\bar B^{0}\big>,
\label{A}
\end{eqnarray}
where $\big<K^{+}K^{-} \bar K^{0}|H^T|\bar B^{0}\big>$ and $\big<K^{+}K^{-} \bar K^{0}|H^P|\bar B^{0}\big>$ refer to the amplitude from the tree and penguin contributions, respectively.
In order to obtain the formalism of the CP asymmetry, we make a definition:
\begin{eqnarray}
A=\big<K^{+}K^{-} \bar K^{0}|H^T|\bar B^{0}\big>[1+re^{i(\delta+\phi)}],
\label{A'}
\end{eqnarray}
where
\begin{eqnarray}
r\equiv\Bigg|\frac{\big<K^{+}K^{-} \bar K^{0}|H^P|\bar B^{0}\big>}{\big<K^{+}K^{-} \bar K^{0}|H^T|\bar B^{0}\big>}\Bigg|.
\label{r}
\end{eqnarray}
The strong phase $\delta$ is generated through Wilson coefficient and the resonance effect in this three-body decay process. The weak phase $\phi$ is associated with the CKM matrix. They can affect the CP asymmetry of this decay process.
Furthermore, we can present the detailed formalisms of the tree and penguin amplitudes by combining above decay diagrams in Fig.1:
\begin{eqnarray}
\big<K^{+}K^{-} \bar K^{0}|H^T|\bar B^{0}\big>=\frac{g_{\rho}}{s_{\rho}}t_{\rho}
+\frac{g_{\phi}}{s_{\phi}s_{\rho}}\widetilde{\Pi}_{\rho\phi}t_{\rho}
+\frac{g_{\omega}}{s_{\omega}s_{\rho}}\widetilde{\Pi}_{\rho\omega}t_{\rho}
+\frac{g_{\omega}}{s_{\omega}}t_{\omega}
+\frac{g_{\phi}}{s_{\phi}s_{\omega}}\widetilde{\Pi}_{\omega\phi}t_{\omega}
+\frac{g_{\rho}}{s_{\rho}s_{\omega}}\widetilde{\Pi}_{\rho\omega}t_{\omega},
\label{Htr}
\end{eqnarray}
\begin{eqnarray}
\begin{split}
\big<K^{+}K^{-} \bar K^{0}|H^P|\bar B^{0}\big>=
&
\frac{g_{\phi}}{s_{\phi}}p_{\phi}
+\frac{g_{\rho}}{s_{\rho}s_{\phi}}\widetilde{\Pi}_{\rho\phi}p_{\phi}
+\frac{g_{\omega}}{s_{\phi}s_{\omega}}\widetilde{\Pi}_{\omega\phi}p_{\phi}
+\frac{g_{\rho}}{s_{\rho}}p_{\rho}
+\frac{g_{\phi}}{s_{\rho}s_{\phi}}\widetilde{\Pi}_{\rho\phi}p_{\rho}
\\
&
+\frac{g_{\omega}}{s_{\rho}s_{\omega}}\widetilde{\Pi}_{\rho\omega}p_{\rho}
+\frac{g_{\omega}}{s_{\omega}}p_{\omega}
+\frac{g_{\phi}}{s_{\omega}s_{\phi}}\widetilde{\Pi}_{\omega\phi}p_{\omega}
+\frac{g_{\rho}}{s_{\rho}s_{\omega}}\widetilde{\Pi}_{\rho\omega}p_{\omega},
\label{Hpe}
\end{split}
\end{eqnarray}
where $t_{\phi}(p_{\phi})$, $t_{\rho}(p_{\rho})$ and $t_{\omega}(p_{\omega})$ are the tree (penguin) contribution for $\bar B^{0}\rightarrow\phi \bar K^{0}$, $\bar B^{0}\rightarrow\rho^{0} \bar K^{0}$ and $\bar B^{0}\rightarrow\omega \bar K^{0}$ decay processes, respectively. The term of $t_{\phi}$ in Eg(7) is ignored since it is zero during our calculation. The coupling constant $g_{V}$ $(V=\phi, \rho, \omega)$ is from the decay process of $V\rightarrow K^{+}K^{-}$. Then, we can obtain
\begin{eqnarray}
\begin{split}
re^{i\delta}e^{i\phi}=
\frac{g_{\omega}p_{\omega}s_{\rho}s_{\phi}+g_{\phi}p_{\phi}s_{\rho}s_{\omega}+
g_{\rho}p_{\rho}s_{\phi}s_{\omega}+g_{\phi}p_{\rho}s_{\omega}\widetilde{\Pi}_{\rho\phi}
+g_{\rho}p_{\phi}s_{\omega}\widetilde{\Pi}_{\rho\phi}}
{g_{\rho}s_{\phi}s_{\omega}t_{\rho}+g_{\omega}s_{\rho}s_{\phi}t_{\omega}+g_{\phi}s_{\omega}t_{\rho}\widetilde{\Pi}_{\rho\phi}
+g_{\rho}s_{\phi}t_{\omega}\widetilde{\Pi}_{\rho\omega}+g_{\phi}s_{\rho}t_{\omega}\widetilde{\Pi}_{\omega\phi}+
g_{\omega}s_{\phi}t_{\rho}\widetilde{\Pi}_{\rho\omega}}
\\
+\frac{g_{\rho}p_{\omega}s_{\phi}\widetilde{\Pi}_{\rho\omega}
+g_{\phi}p_{\omega}s_{\rho}\widetilde{\Pi}_{\omega\phi}+g_{\omega}p_{\rho}s_{\phi}\widetilde{\Pi}_{\rho\omega}
+g_{\omega}p_{\phi}s_{\rho}\widetilde{\Pi}_{\omega\phi}}
{g_{\rho}s_{\phi}s_{\omega}t_{\rho}+g_{\omega}s_{\rho}s_{\phi}t_{\omega}+g_{\phi}s_{\omega}t_{\rho}\widetilde{\Pi}_{\rho\phi}
+g_{\rho}s_{\phi}t_{\omega}\widetilde{\Pi}_{\rho\omega}+g_{\phi}s_{\rho}t_{\omega}\widetilde{\Pi}_{\omega\phi}+
g_{\omega}s_{\phi}t_{\rho}\widetilde{\Pi}_{\rho\omega}}.
\label{rdtdirive}
\end{split}
\end{eqnarray}

Define
\begin{eqnarray}
\frac{p_{\omega}}{t_{\rho}}\equiv r_{1}
e^{i(\delta_\lambda+\phi)},
\quad\frac{p_{\phi}}{t_{\rho}}\equiv r_{2}e^{i(\delta_\chi+\phi)},
\quad\frac{t_{\omega}}{t_{\rho}}\equiv r_{3} e^{i\delta_\alpha},
\quad\frac{p_{\rho}}{p_{\omega}}\equiv r_{4} e^{i\delta_\beta}, \label{def}
\end{eqnarray}
where $\delta_\lambda$, $\delta_\chi$, $\delta_\alpha$ and $\delta_\beta$ are strong phases. Then we can take them into the Eq.(\ref{rdtdirive}) and make some simplification:
\begin{eqnarray}
\begin{split}
re^{i\delta}=
&
\frac{r_{1}e^{i\delta_\lambda}s_{\phi}s_{\rho}g_{\omega}
+r_{2}e^{i\delta_\chi}s_{\omega}s_{\rho}g_{\phi}
+r_{1}e^{i\delta_\lambda}r_{4}e^{i\delta_\beta}s_{\omega}s_{\phi}g_{\rho}
+r_{2}e^{i\delta_\chi}s_{\omega}g_{\rho}\widetilde{\Pi}_{\rho\phi}
+r_{1}e^{i\delta_\lambda}r_{4}e^{i\delta_\beta}s_{\omega}g_{\phi}\widetilde{\Pi}_{\rho\phi}}
{r_{3}e^{i\delta_\alpha}s_{\phi}s_{\rho}g_{\omega}+s_{\phi}s_{\omega}g_{\rho}
+s_{\omega}g_{\phi}\widetilde{\Pi}_{\rho\phi}
+r_{3}e^{i\delta_\alpha}s_{\phi}g_{\rho}\widetilde{\Pi}_{\rho\omega}
+r_{3}e^{i\delta_\alpha}s_{\rho}g_{\phi}\widetilde{\Pi}_{\omega\phi}
+s_{\phi}g_{\omega}\widetilde{\Pi}_{\rho\omega}}\\
&
+\frac{r_{1}e^{i\delta_\lambda}s_{\phi}g_{\rho}\widetilde{\Pi}_{\rho\omega}
+r_{1}e^{i\delta_\lambda}s_{\rho}g_{\phi}\widetilde{\Pi}_{\omega\phi}
+r_{1}e^{i\delta_\lambda}r_{4}e^{i\delta_\beta}s_{\phi}g_{\omega}\widetilde{\Pi}_{\rho\omega}
+r_{2}e^{i\delta_\chi}s_{\rho}g_{\omega}\widetilde{\Pi}_{\omega\phi}}
{r_{3}e^{i\delta_\alpha}s_{\phi}s_{\rho}g_{\omega}+s_{\phi}s_{\omega}g_{\rho}
+s_{\omega}g_{\phi}\widetilde{\Pi}_{\rho\phi}
+r_{3}e^{i\delta_\alpha}s_{\phi}g_{\rho}\widetilde{\Pi}_{\rho\omega}
+r_{3}e^{i\delta_\alpha}s_{\rho}g_{\phi}\widetilde{\Pi}_{\omega\phi}
+s_{\phi}g_{\omega}\widetilde{\Pi}_{\rho\omega}},
\end{split}
\end{eqnarray}

The $\phi$ is related to $\frac{V_{tb}V_{ts}^{*}}{V_{ub}V_{us}^{*}}$ from CKM matrix. Therefore, we can get ${\rm sin}\phi =-\frac{\eta}{\sqrt{\rho^2+\eta^2}}$ and
${\rm cos}\phi = -\frac{\rho}{\sqrt{\rho^2+\eta^2}}$ through the Wolfenstein parameters.
Thus we can make the definition about the CP asymmetry:
\begin{eqnarray}
\begin{split}
A_{cp} = \frac{\left| A \right|^2-\left| \overline{A} \right|^2}{\left| A \right|^2+\left| \overline{A} \right|^2}= \frac{-2r{\rm sin\delta}{\rm sin\phi}}{1+2r{\rm cos\delta}{\rm cos\phi}+r^2}.
\end{split}
\label{cp-define}
\end{eqnarray}

\subsection{\label{subsec:form} The form of localised CP asymmetry}
In previous experimental measurements, large localised CP asymmetry has been observed from  the decay mode of $B^{\pm}\rightarrow \phi K^{\pm}\rightarrow K^{+}K^{-}K^{\pm}$ in the region of $m_{K^+ K^-}$$<$1.04 \textrm{GeV} and the decay mode of $B^{\pm}\rightarrow \phi \pi^{\pm}\rightarrow K^{+}K^{-}\pi^{\pm}$ in the region of $m^2_{K^+ K^-}$$<$1.5 $\textrm{GeV}^2$ under the range of $\phi$ \cite{R2013,PL2014,B2012}. For simplify, we take the $\bar B^{0}\rightarrow \phi(\omega,\rho^0) \bar K^{0}\rightarrow K^{+}K^{-} \bar K^{0}$ decay channel as an example to introduce the localised CP asymmetry in the certain energy region by integrating the $A_{CP}$. First, the total amplitude of $\bar B^{0}\rightarrow K^{+}K^{-} \bar K^{0}$ can be regarded as two components. We can get their forms in the decay process of $\bar B^{0}\rightarrow\phi \bar K^{0}\rightarrow K^{+}K^{-} \bar K^{0}$ as following:
\begin{equation}
M_{\bar B^{0}\rightarrow \phi \bar K^{0}}^{\lambda}=\alpha p_{\bar B^{0}} \cdot \epsilon^{*}(\lambda),
\end{equation}
\begin{equation}
M_{\phi \rightarrow K^{+} K^{-}}^{\lambda}=g_{\phi}\epsilon(\lambda)\left(p_{1}-p_{2}\right),
\end{equation}
where $\epsilon$ is the polarization vector of $\phi$. $\lambda$ represents the direction of polarization for $\epsilon$. $p_{\bar B^{0}}$ is the momenta of $\bar B^{0}$ meson. $\alpha$ refers to the part of the amplitude which is independent of $\lambda$. For the decay process $\phi\rightarrow K^{+}K^{-}$, $p_{1}$ and $p_{2}$ are the momentum of $K^{+}$ and $K^{-}$ produced by $\phi$. $g_{\phi}$ is introduced to the effective coupling constant. $s_{\phi}$ refers to the gluon propagator. Therefore the total amplitude of $\bar B^{0}\rightarrow\phi \bar K^{0}\rightarrow K^{+}K^{-} \bar K^{0}$ decay process can be expressed as:
\begin{equation}
\begin{aligned}
A &=\alpha p_{\bar B^{0}}^{\mu} \frac{\sum_{\lambda} \epsilon_{\mu}^{*}(\lambda) \epsilon_{\nu}(\lambda)}{s_{\phi}} \frac{g_{\phi}}{s_{\phi}}\left(p_{1}-p_{2}\right)^{\nu} \\
&=-\frac{g_{\phi} \alpha}{s_{\phi}} \cdot p_{\bar B^{0}}^{\mu}\left[g_{\mu \nu}-\frac{\left(p_{1}+p_{2}\right)_{\mu}\left(p_{1}+p_{2}\right)_{\nu}}{m_{\phi}^{2}}\right]\left(p_{1}-p_{2}\right)^{\nu},
\end{aligned}
\end{equation}
where $\sqrt{s}$ is the low invariant mass of the $K^{+}K^{-}$ pair. If $\sqrt{s^{\prime}}$ is the high invariance mass of the $ K^{+} K^{-}$ pair, we make $ s_{\max }^{\prime}$ be the maximum values of $s^{\prime}$ for a fixed $s$ and $s_{\min }^{\prime}$ be minimum values of $s^{\prime}$ for a fixed $s$ \cite{R2014}. we get $m^{2}_{ij}$=$p^{2}_{ij}$ by conservation of energy and momentum in the three body decay process. Thus, we make the amplitude as the following expression:
\begin{equation}
A =\frac{g_{\phi}}{s_{\phi}} \cdot \frac{M_{\bar B^{0}\rightarrow \phi \pi^{0}( \bar K^{0})}^{\lambda}}{p_{\bar B^{0}} \cdot \epsilon^{*}} \cdot\left(\Sigma-s^{\prime}\right)=\left(\Sigma-s^{\prime}\right)\cdot \mathcal{N},
\end{equation}

We can integrate the denominator and numerator of $A_{CP}$ within the range of $\Omega \left(s_{1}<s<s_{2}, s_{1}^{\prime}<s^{\prime}< s_{2}^{\prime}\right)$. $\mathcal{N}$ is the substitution of the previous formula. $\Sigma=\frac{1}{2}\left(s_{\max }^{\prime}+s_{\min }^{\prime}\right)$ is related to $s$. Since $s$ varies in a small region, therefore $\Sigma$ can be treated as a constant approximately. Thus we can cancel the result of $\int_{s_{1}^{'}}^{s_{2}^{'}}{ds^{'}}\left( \Sigma -s^{'} \right) ^2$. We obtain the localized integrated CP asymmetry, which takes this form \cite{WZWG2015}:
\begin{equation}
A_{CP}^{\Omega}=\frac{\int_{s_{1}}^{s_{2}} \mathrm{~d} s \int_{s_{1}^{\prime}}^{s_{2}^{\prime}} \mathrm{d} s^{\prime}\left(\Sigma-s^{\prime}\right)^{2}\left(|\mathcal{N}|^{2}-|\overline{\mathcal{N}}|^{2}\right)}{\int_{s_{1}}^{s_{2}} \mathrm{~d} s \int_{s_{1}^{\prime}}^{s_{2}^{\prime}} \mathrm{d} s^{\prime}\left(\Sigma-s^{\prime}\right)^{2}\left(|\mathcal{N}|^{2}+|\overline{\mathcal{N}}|^{2}\right)}.
\end{equation}

We consider that $A_{CP}^{\varOmega}$ is independent of the high invariant mass of positive and negative meson pairs in this way. In addition, we also consider the $s$ dependence between the values of $s_{\max}^{'}$ and $s_{\min}^{'}$ in our calculations. It is assumed that $s_{\min}^{'}<s^{'}<s_{\max}^{'}$ represents an integral interval of the high invariance mass of $K ^+K ^-$, while $\int_{s_{\min}^{'}}^{s_{\max}^{'}}{ds^{'}}\left( \Sigma -s^{'} \right) ^2$ represents the factor that is dependent upon $s$. Therefore the terms $\int_{s_{1}^{\prime}}^{s_{2}^{\prime}} \mathrm{d}s^{\prime}\left(\Sigma-s^{\prime}\right)$ are ignored, and $A_{CP}^{\Omega}$ becomes independent of the high invariance mass of $K^{+}K^{-}$.

\section{\label{sec:cpv1}The calculation of decay Amplitude}
We calculate the CP asymmetry by taking the method of the quasi-two-body decay process, where the contributions from the tree-level and penguin-level are presented. In the two-body decay of the B meson, the form factor of the B to the final hadron transition is mainly the contribution of the non-perturbative region. The non-factorization effect of the hadron matrix element is mainly the exchange of hard gluons. We use the QCD factorization approach to calculate the quasi-two-body decay amplitudes. We have also finished the calculation of the associated hard-scattering kernels. In the phenomenological analysis, we calculate the ``chirality-enhancement" term and the weak annihilation contributions \cite{N2003}. Based on CKM matrix elements of $V_{ub}V^{*}_{us}$ and $V_{tb}V^{*}_{ts}$, the decay amplitude of $\bar{B}^{0}\rightarrow \phi(\rho^0,\omega) \bar K^{0}\rightarrow K^+ K^-\bar K^{0}$ in QCDF approach can be written as:
\begin{eqnarray}
\begin{array}{c}
	A\left(\bar{B}^{0}\rightarrow \phi \left( \phi \rightarrow K^+ K^- \right) \bar K^{0} \right) ={\sum_{\lambda =0,\pm 1}}{\frac{G_FP_{\bar{B}^{0}}\cdot \epsilon ^*\left( \lambda \right) g_{\phi}\epsilon \left( \lambda \right) \cdot \left( p_{K ^+}-p_{K ^-} \right)}{s_{\phi}}}\\
	\\
	\times \left\{V_{tb}V_{ts}^{*} \left[-\sqrt{2}m_{\phi} (\epsilon \cdot p_{K})f_{\phi}F_{1}^{B\rightarrow K}\left( a_3+a_5+a_4-\frac{1}{2}a_{10}\right.\right.\right. \\
	\\
\left.\left.\left.-\frac{1}{2}a_7-\frac{1}{2}a_9\right)-\frac{\sqrt{2}}{2}f_B f_K f_{\phi}b_{3}(\phi,K)-\frac{\sqrt{2}}{4}f_B f_K f_{\phi}b_{3}^{e\omega}(\phi,K)\right]\right\},
\end{array}
\end{eqnarray}

\begin{eqnarray}
\begin{array}{c}
	A\left(\bar{B}^{0}\rightarrow \rho^0 \left(\rho^0 \rightarrow K^+ K^- \right) \bar K^{0} \right) ={\sum_{\lambda =0,\pm 1}}{\frac{G_FP_{\bar{B}^{0}}\cdot \epsilon ^*\left( \lambda \right) g_{\rho}\epsilon \left( \lambda \right) \cdot \left( p_{K ^+}-p_{K ^-} \right)}{s_{\rho}}}\\
	\\
	\times \left\{V_{ub}V_{us}^{*}m_{\rho}(\epsilon \cdot p_{K})f_{\rho}F_{1}^{B\rightarrow K}a_2+V_{tb}V_{ts}^{*}\left[m_{\rho}(\epsilon \cdot p_{K})f_{K} \right.\right. \\
	\\
\left.\left. A_{0}^{B\rightarrow \rho}\left( a_4-\frac{1}{2}a_{10}+a_6Q -\frac{1}{2}a_8Q\right)-m_{\rho}(\epsilon \cdot p_{K})f_{\rho}F_{1}^{B\rightarrow K}\right.\right.\\
	\\
\left.\left.(\frac{3}{2}a_{7}+\frac{3}{2}a_9)+\frac{1}{2}f_B f_K f_{\rho} b_{3}(K,\rho)-\frac{1}{4}f_B f_K f_{\rho}b_{3}^{e\omega}(K,\rho)\right]\right\},
\end{array}
\end{eqnarray}

\begin{eqnarray}
\begin{array}{c}
	A\left(\bar{B}^{0}\rightarrow \omega \left(\omega \rightarrow K^+ K^- \right) \bar K^{0} \right) ={\sum_{\lambda =0,\pm 1}}{\frac{G_FP_{\bar{B}^{0}}\cdot \epsilon ^*\left( \lambda \right) g_{\omega}\epsilon \left( \lambda \right) \cdot \left( p_{K ^+}-p_{K ^-} \right)}{s_{\omega}}}\\
	\\
	\times \left\{V_{ub}V_{us}^{*}m_{\omega} (\epsilon \cdot p_{K})f_{\omega}F_{1}^{B\rightarrow K}a_2+V_{tb}V_{ts}^{*}\left[-m_{\omega} (\epsilon \cdot p_{K})f_{K}\right.\right. \\
	\\
\left.\left.A_{0}^{B\rightarrow \omega}\left(a_4-\frac{1}{2}a_{10}+a_6Q-\frac{1}{2}a_8Q \right)-m_{\omega} (\epsilon \cdot p_{K})f_{\omega}F_{1}^{B\rightarrow K}\left(2a_3\right.\right.\right. \\
	\\
\left.\left.\left.+2a_5+\frac{1}{2}a_{7}+\frac{1}{2}a_9\right)-\frac{1}{2}f_B f_K f_{\omega}b_{3}(K,\omega)+\frac{1}{4}f_B f_K f_{\omega}b_{3}^{e\omega}(K,\omega)\right]\right\}.
\end{array}
\end{eqnarray}
where $f_{\pi}$, $f_{B}$, $f_{\phi(\rho, \omega)}$ represent the decay constants and $a_{n}$(n=1,2,3...) is related to the Wilson coefficient $C_{n}$ under the QCD factorization theory \cite{PRD2000}. $A_{0}^{B\rightarrow\rho}$, $A_{0}^{B\rightarrow\omega}$ and $F_{1}^{B\rightarrow\pi}$ are the form factor from the non-perturbative contribution. Besides, $b_{1(\rho,\pi)}$ and $b_{3,4}^{e\omega}(\rho,\pi)$ are the contribution from the annihilation process \cite{J2016}. $\epsilon$ and $p_{K}$ represent the polarization vector and momentum of K meson. In order to simplify the calculation, we define the operator Q ($Q=\frac{-2m_{K^{0}}^{2}}{(m_b+m_d)(m_d+m_s)}$) \cite{PR2000,PB2001}.

For the hard scattering process of the spectator quark, the end point integrals of these logarithmic divergence are expressed by phenomenological parameters. The contribution of the twist-3 distribution amplitude of the final pseudo-scalar meson is considered to investigate the annihilation amplitude \cite{PB2001}. In the annihilation decay process, $b$ represents the annihilation coefficient where $b_{1,2}$,
$b_{3,4}$ and $b_{3,4}^{e\omega}$ corresponds to the effective operator $Q_{1,2}$, QCD penguin operator $Q_{3-6}$ and the weak electric penguin operator $Q_{7-10}$, respectively \cite{10M2003}.
The amplitude is parameterized depending on the contributions of tree-level and penguin-level, where the parameters $X_H$ and $X_A$ are introduced. We define $X_{H}=\int_{0}^{1} \frac{d{y}}{1-y}$ and $X_{A}=\int_{0}^{1} \frac{d{x}}{x}$ to deal with the contribution from the annihilation process, which can not be ignored \cite{3Li2019}. We also consider the hard scattering of the spectator quark and the contribution from the annihilation process, which provide the information about the strong phase.

\section{\label{sum}Numerical results}
\subsection{\label{subsec:form}The results of CP asymmetry}

\begin{figure}[!htbp]
	\centering
	\begin{minipage}[h]{0.45\textwidth}
		\centering
		\includegraphics[height=4cm,width=6.5cm]{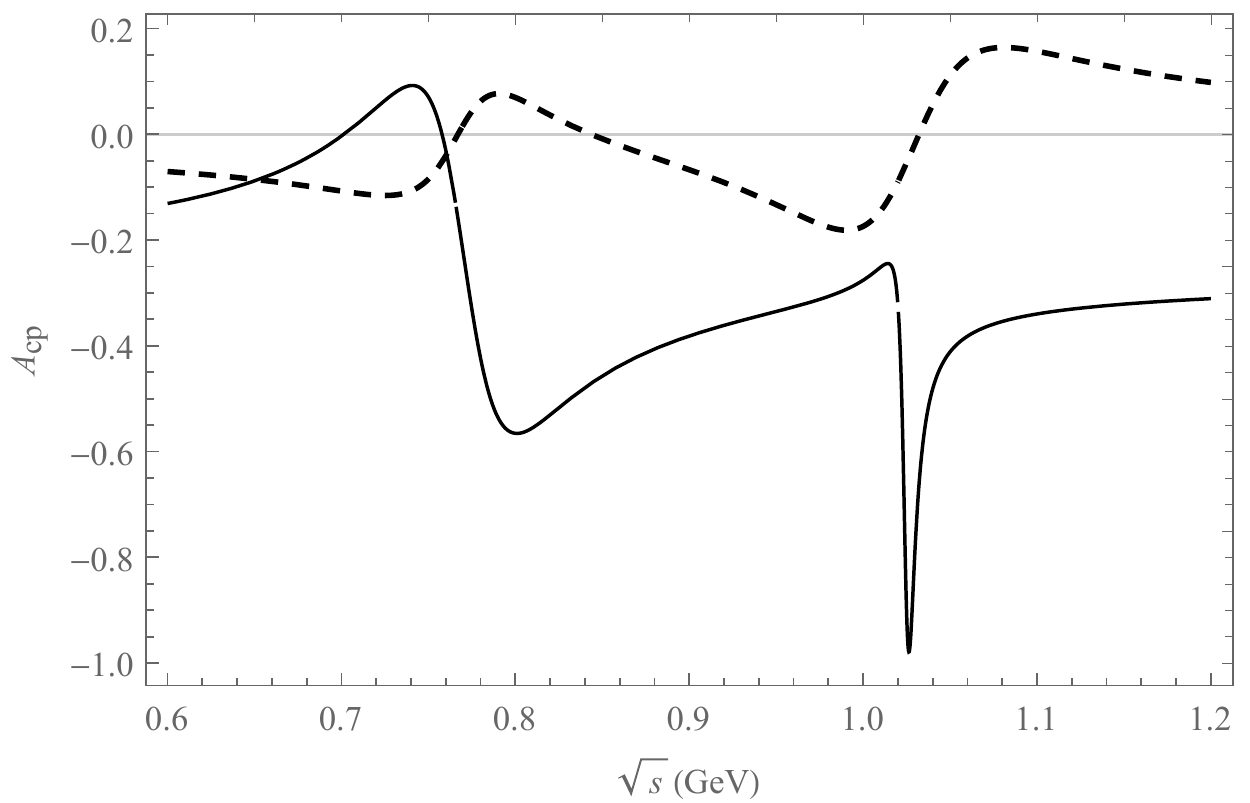}
		\caption{Plot of $A_{CP}$ as a function of $\sqrt{s}$, the solid line corresponds to the decay channel of $\bar B^{0}\rightarrow K^{+}K^{-}\pi^{0}$ and the dotted line refers to the decay channel of $\bar B^{0}\rightarrow K^{+}K^{-} \bar K^{0}$.}
		\label{fig2}
	\end{minipage}
	\quad
	\begin{minipage}[h]{0.45\textwidth}
		\centering
		\includegraphics[height=4cm,width=6.5cm]{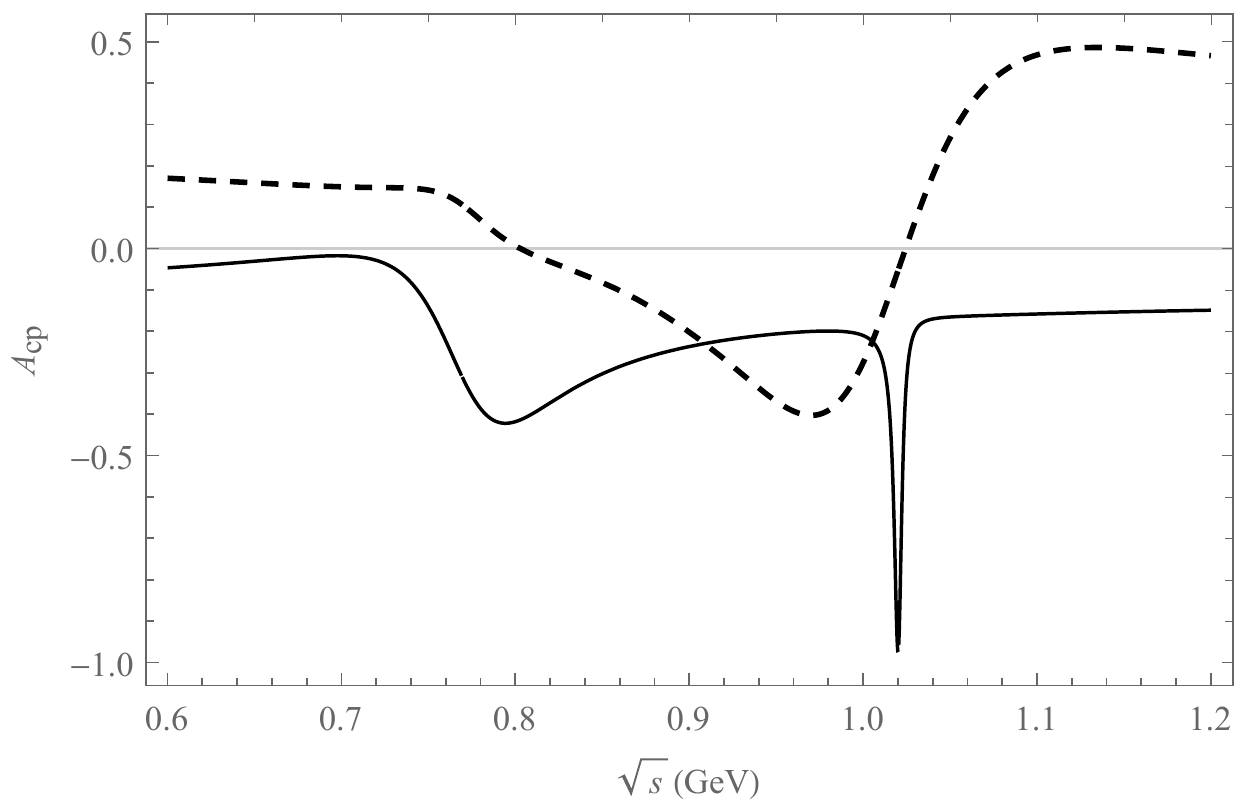}
		\caption{Plot of $A_{CP}$ as a function of $\sqrt{s}$, the solid line corresponds to the decay channel of $B^{-}\rightarrow K^{+}K^{-}\pi^{-}$ and the dotted line refers to the decay channel of $B^{-}\rightarrow K^{+}K^{-}K^{-}$.}
		\label{fig3}
	\end{minipage}
\end{figure}

\begin{figure}[!htbp]
\centering
\begin{minipage}[h]{0.45\textwidth}
\centering
\includegraphics[height=4cm,width=6.5cm]{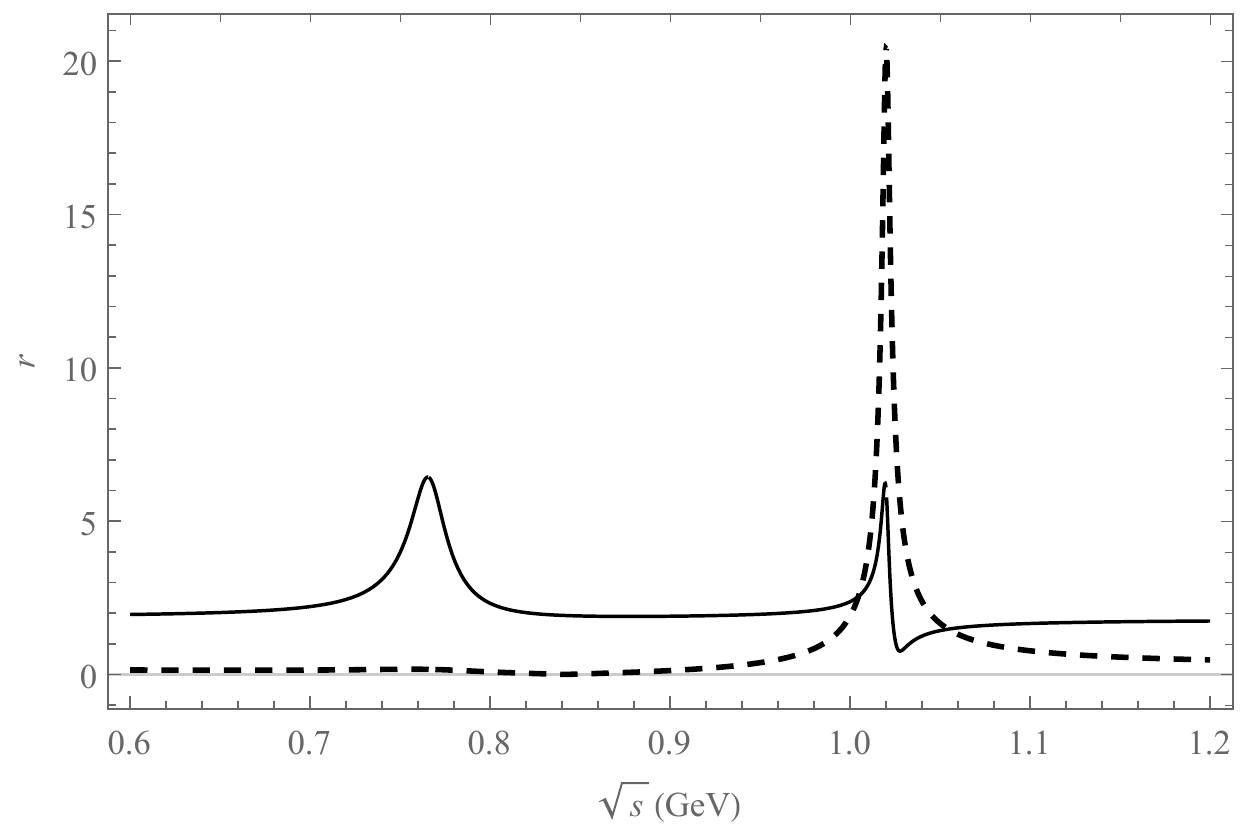}
\caption{Plot of $r$ as a function of $\sqrt{s}$, the solid line corresponds to the decay channel of $\bar B^{0}\rightarrow K^{+}K^{-}\pi^{0}$ and the dotted line refers to the decay channel of $\bar B^{0}\rightarrow K^{+}K^{-} \bar K^{0}$.}
\label{fig4}
\end{minipage}
\quad
\begin{minipage}[h]{0.45\textwidth}
\centering
\includegraphics[height=4cm,width=6.5cm]{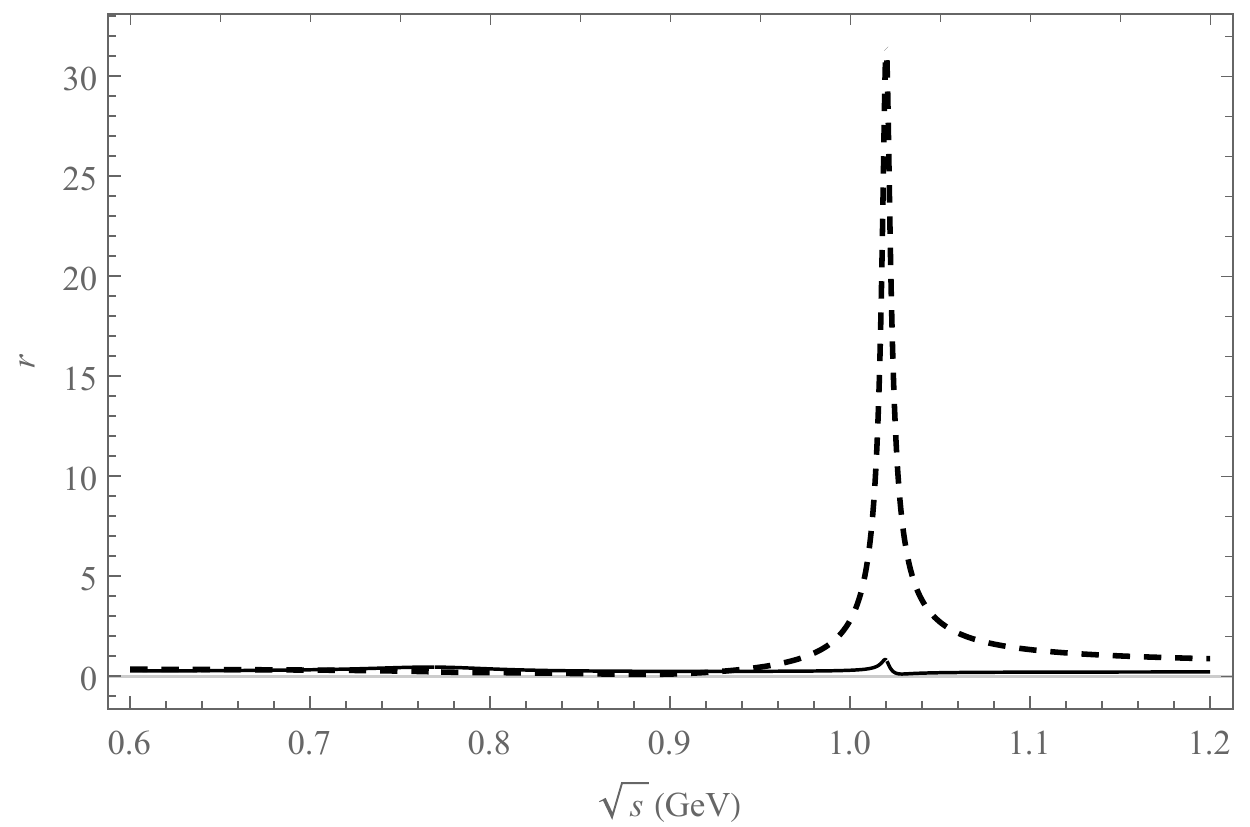}
\caption{Plot of $r$ as a function of $\sqrt{s}$, the solid line corresponds to the decay channel of $B^{-}\rightarrow K^{+}K^{-}\pi^{-}$ and the dotted line refers to the decay channel of $B^{-}\rightarrow K^{+}K^{-}K^{-}$.}
\label{fig5}
\end{minipage}
\end{figure}

\begin{figure}[!htbp]
\centering
\begin{minipage}[h]{0.45\textwidth}
\centering
\includegraphics[height=4cm,width=6.5cm]{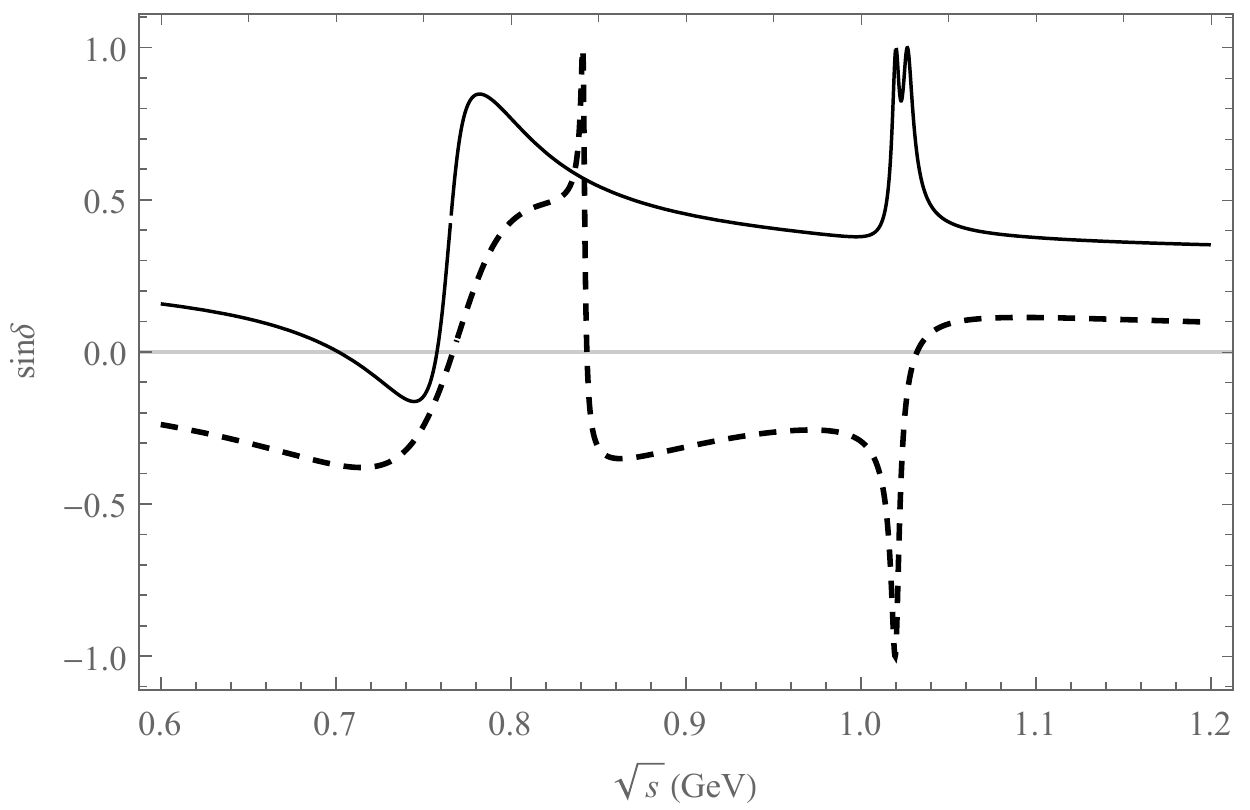}
\caption{Plot of ${\rm{sin}}\delta$ as a function of $\sqrt{s}$, the solid line is the decay channel of $\bar B^{0}\rightarrow K^{+}K^{-}\pi^{0}$ and the dotted line refers to the decay channel of $\bar B^{0}\rightarrow K^{+}K^{-} \bar K^{0}$.}
\label{fig6}
\end{minipage}
\quad
\begin{minipage}[h]{0.45\textwidth}
\centering
\includegraphics[height=4cm,width=6.5cm]{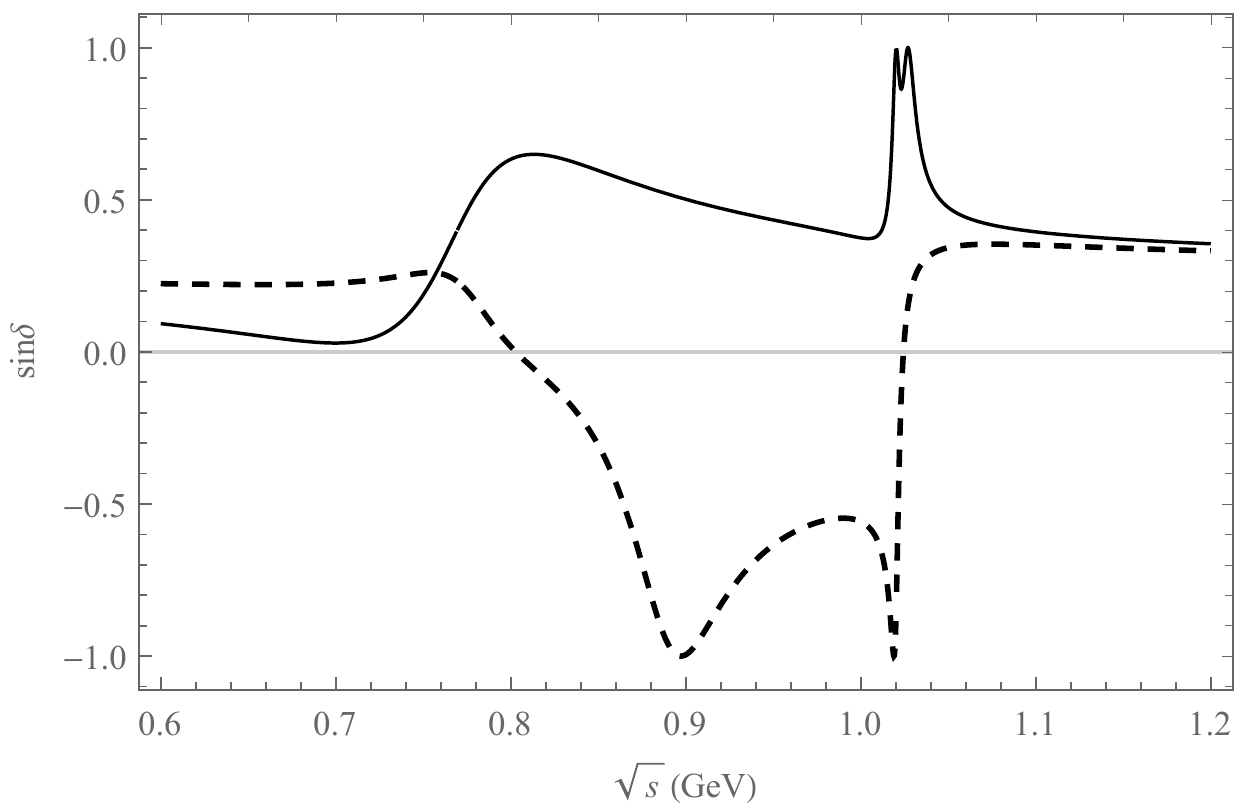}
\caption{Plot of ${\rm{sin}}\delta$ as a function of $\sqrt{s}$, the solid line is the decay channel of $B^{-}\rightarrow K^{+}K^{-}\pi^{-}$ and the dotted line refers to the decay channel of $B^{-}\rightarrow K^{+}K^{-}K^{-}$.}
\label{fig7}
\end{minipage}
\end{figure}

The numerical results clarify the relationship between the CP asymmetry ($r$ and strong phase $\delta$) and $\sqrt{s}$ under the resonance effects of the three vector mesons, which are shown in Fig.\ref{fig2} to Fig.\ref{fig7}. In order to better compare with the experiment data, we consider the region of $m_{K^{+}K^{-}}$$<$1.04 $\textrm{GeV}$ which is the experimental result \cite{B2012}. The energy region both low and high limits by up to 0.5 $\textrm{GeV}^2$ for $\phi(1020)$ in $B^{\pm}\rightarrow K^{+}K^{-}K^{\pm}$ decay, and by up to 1 $\textrm{GeV}^2$ for all other decays in LHC \cite{LHC2022}. We choose the region of 0.6 $\textrm{GeV}$-1.2 $\textrm{GeV}$ under our theoretical framework which is the main region of resonance for the decay process of $B\rightarrow K^{+}K^{-}\pi(K)$ to draw the figures. One can find that the CP asymmetry change sharply for the decay process of $\bar{B}^{0}\rightarrow K^{+}K^{-}\bar K^{0}(\pi^{0})$ and $B^{-}\rightarrow K^{+}K^{-}K^{-}(\pi^{-})$ from the $\phi\rightarrow K^{+}K^{-}$, $\rho\rightarrow K^{+}K^{-}$, $\omega\rightarrow K^{+}K^{-}$ resonance in Fig.\ref{fig2} and Fig.\ref{fig3} where $\phi$ is the main contribution.

For the decay mode of $\bar{B}^{0}\rightarrow K^{+}K^{-}\pi^{0}$, we find there are large CP asymmetries, which vary from $9.58\%$ to $-56.2\%$ at the $\rho$ and $\omega$ resonance range and from $-24.5\%$ to $-98.5\%$ at the $\phi$ resonance range in Fig.\ref{fig2}. In addition, the CP asymmetry change lightly from the $\bar{B}^{0}\rightarrow K^{+}K^{-} \bar K^{0}$ decay process comparing with the process of $\bar{B}^{0}\rightarrow K^{+}K^{-}\pi^{0}$. The CP asymmetries change from $7.74\%$ to $-11.8\%$ at the $\rho$ and $\omega$ resonance range and vary from $16.2\%$ to $-18.3\%$ at the $\phi$ resonance range.

The maximum value of the CP asymmetries of the decay channel of $B^{-}\rightarrow K^{+}K^{-}\pi^{-}$ can reach $-45.3\%$ at the $\rho$ and $\omega$ resonance range and $-97.7\%$ at the $\phi$ resonance range in Fig.\ref{fig3}. Besides, we obtain the CP asymmetries vary from $13.75\%$ to $11.2\%$ at the $\rho$ and $\omega$ resonance range and from $46.8\%$ to $-40.7\%$ at the $\phi$ resonance range for the decay channel of $B^{-}\rightarrow K^{+}K^{-}K^{-}$.

The decay channel of $\bar{B}^{0}\rightarrow K^{+}K^{-} \bar K^{0}$ is similarly to the decay channel of $B^{-}\rightarrow K^{+}K^{-}K^{-}$ which the CP asymmetry changes little relatively in the resonance of $\rho$ and $\omega$ mass and mainly appear in the resonance of $\phi$ mass. Obviously, one can find that the contribution of the resonance effect is obvious around the range of the $\phi$ mass when the final state is consisted of K meson, which can be seen in the Fig.2 and Fig.3. But the mechanism of resonance has little change near the range of $\rho$ mass and $\omega$ mass for the decay channel of $\bar{B}^{0}\rightarrow K^{+}K^{-} \bar K^{0}$.

The ratio $r$ of the penguin-level contribution and tree-level contribution affects the CP asymmetry from Eq.(\ref{cp-define}). Then, we present the relationship between the $r$ and $\sqrt{s}$ by taking the central parameter values of CKM matrix elements in Fig.\ref{fig4} and Fig.\ref{fig5}. It is obvious that the rangeability of $r$ in the $\phi$ resonance range is large compared with the resonance range of $\rho$ and $\omega$ except the decay mode of $\bar{B}^{0}\rightarrow K^{+}K^{-}\pi^{0}$. We also find that the value of $r$ changes obviously at the region of the $\phi$ mass when the final state is consisted of K meson from the decay modes of $\bar B^{0}\rightarrow K^{+}K^{-} \bar K^{0}$ and $B^{-}\rightarrow K^{+}K^{-}K^{-}$.

The CP asymmetry also depends on the strong phase due to Eq.(\ref{cp-define}).
We express the plots between ${\rm{sin}}\delta$ and $\sqrt{s}$ in Fig.\ref{fig6} and Fig.\ref{fig7}. Compared with the above plots which the main contribution is from the range of $\phi$, the Fig.6 and Fig.7 show the large CP asymmetry can also occur in the ranges of $\omega$ and $\rho$. Therefore, the resonance of the vector mesons can affect the CP asymmetry of the three-body decay process.

\subsection{\label{subsec:form}The value of localised CP asymmetry}

\begin{table}[!ht]
\renewcommand
\arraystretch {3}
\centering %
\renewcommand{\arraystretch}{2.0} %
\caption{The comparison of $A_{cp}$ in our work with experiments without $\phi-\rho-\omega$ mixing}
\setlength{\tabcolsep}{3.5mm}
\begin{tabular}{c|c|c|c}
\hline
Decay channel
& The experiment result
& {\makecell[c]{$\phi-\rho-\omega$ mixing\\(0.6-1.2 \textrm{GeV})}}
& {\makecell[c]{$\phi-\rho-\omega$ mixing\\(0.9-1.1 \textrm{GeV})}}
\\ \hline  $\bar B^{0}\to\phi\pi^{0}\to K^{+}K^{-}\pi^{0}$   & -  & -0.238$\pm$0.013$\pm$0.002 & -0.337$\pm$0.004$\pm$0.002
\\ \hline  $\bar B^{0}\to\phi K^{0}\to K^{+}K^{-}\bar K^{0}$   & {\makecell[c]{-0.08$\pm$0.18$\pm$0.04 BaBar \cite{B2007}\\($m_{K^+ K^-}$$<$1.1 $\textrm{GeV}$)}}  & -0.074$\pm$0.005$\pm$0.001  & -0.084$\pm$0.001$\pm$0.007
\\ \hline  $B^{-}\to\phi\pi^{-}\to K^{+}K^{-}\pi^{-}$   & {\makecell[c]{-0.141$\pm$0.040$\pm$0.018 LHC \cite{PL2014}\\($m^2_{K^+ K^-}$$<$1.5 $\textrm{GeV}^2$)}} & -0.180$\pm$0.007$\pm$0.003 &-0.227$\pm$0.002$\pm$0.005
\\ \hline  $B^{-}\to\phi K^{-}\to K^{+}K^{-}K^{-}$   & {\makecell[c]{{\makecell[c]{0.128$\pm$0.044$\pm$0.013 BaBar\cite{B2012}\\($m_{K^+ K^-}$$<$1.04 $\textrm{GeV}$)}}\\{\makecell[c]{-0.004$\pm$0.010$\pm$0.007 LHC \cite{LHC2022}\\($m^2_{K^+ K^-}$$<$1 $\textrm{GeV}^2$)}}}}  & -0.034$\pm$0.015$\pm$0.002 &-0.048$\pm$0.018$\pm$0.001
\\ \hline
\end{tabular}
\end{table}

In our work, we integrate the CP asymmetry for the processes of $\bar B^{0}\rightarrow \phi \bar K^{0}(\pi^{0}) \rightarrow K^{+}K^{-}\bar K^{0}(\pi^{0})$ and $B^{-}\rightarrow \phi K^{-}(\pi^{-}) \rightarrow K^{+}K^{-}K^{-}(\pi^{-})$ when the invariant masses of $m_{K^{+}K^{-}}$ are in the region of 0.6 $\textrm{GeV}$-1.2 $\textrm{GeV}$ from the resonance range of $\phi$, $\rho$ and $\omega$ and 0.9 $\textrm{GeV}$-1.1 $\textrm{GeV}$ around the $\phi$ mass range.
The results are also presented in Table I which indicates the energy intervals of the experimental results. One can easily find that the mixing mechanism changes the values of CP asymmetry comparing to the experimental results without mixing for some decay channels from the resonant ranges. Especially for the $B^{-}\to\phi \pi^{-}(K^{-})\to K^{+}K^{-}\pi^{-}(K^{-})$, the CP asymmetry under the $\phi-\rho-\omega$ mixing is obvious relatively. The central value of CP asymmetry from $\phi$, $\rho$ and $\omega$ mixing can increase $60\%$ as much as the non-resonant result of experiments for the decay mode
$B^{-}\to\phi\pi^{-}\to K^{+}K^{-}\pi^{-}$ around 0.9 \textrm{GeV}-1.1 \textrm{GeV}.
In the ranges of 0.6 \textrm{GeV} to 1.2 \textrm{GeV} and 0.9 \textrm{GeV}-1.1 \textrm{GeV}, the central values are enhanced large than that of non-mixing experimental results from the LHC expriments for the decay channel $B^{-}\to\phi K^{-}\to K^{+}K^{-}K^{-}$.

The measurement of CP asymmetry in the decay of B meson have become more accurate due to the large number of data collected by experiments in recent years. The CP asymmetry from the decay mode of $\bar B^{0}\rightarrow \phi \bar K^{0}\rightarrow K^{+}K^{-}\bar K^{0}$ has been presented by BarBar \cite{B2007}, which describes the analysis of time-dependent Dalitz plot and extracts the values of the CP asymmetry by taking into account the complex amplitudes describing the entire $B^{0}$ and $\bar B^{0}$ Dalitz plots. Besides, the CP asymmetry from the decay mode of $B^{-}\rightarrow \phi K^{-}(\pi^{-}) \rightarrow K^{+}K^{-}K^{-}(\pi^{-})$ has been presented by LHC \cite{PL2014,R2013,R2020prl,LHC2022} and BarBar \cite{B2012}. The results from these decay processes are compared with our work in Table I.

As shown in the Table I, the decay mode of ${B}^{-}\to\phi K^{-}(\pi)^{-}\rightarrow K^{+}K^{-}K^{-}(\pi)^{-}$ has been measured by LHC and BaBar experiments \cite{PL2014,B2012,LHC2022}. The experimental result where the first uncertainty is from statistics, the second is the experimental systematic, and the third is due to the CP
asymmetry of the $B^{\pm}\to$$J$$/$$\psi$$K^{\pm}$ reference mode. In our work under the resonance mechanism, the first uncertainty is statistics, the second uncertainty is the form factor and Wolfstein parameters, and the third is due to the phenomenological parameter to calculate the decay amplitude in the QCDF method.

\section{\label{sum}SUMMARY AND CONCLUSION}
It is shown that CP asymmetry is affected via the mixing of $V\rightarrow K^{+}K^{-}$ $(V=\phi, \rho, \omega)$ from the decay modes of $B \rightarrow KK\pi(K)$ when the invariant masses of $K^{+}K^{-}$ pairs are near the $\phi$, $\rho$ and $\omega$ resonance ranges in QCD factorization.
The value of CP asymmetry are enhanced than that of non-mixing experimental results for the decay mode of $B^{-}\to\phi\pi^{-}(K^{-})\to K^{+}K^{-}\pi^{-}(K^{-})$ in the resonance range. The interferences of $\phi-\omega$, $\phi-\rho$ and $\omega-\rho$ leads to the new generation of strong phase to affect the CP asymmetry from the decay modes of $B \rightarrow KK\pi(K)$.

We also calculte the CP asymmetry from the non-interference of $\phi$, $\rho$ and $\omega$ for the decay mode of $B^{-}\rightarrow \phi \pi^{-}\rightarrow K^{+}K^{-}\pi^{-}$ for comparing the results of experiments. The CP asymmetry of $B^{-}\rightarrow\phi\pi^{-}\rightarrow K^{+}K^{-}\pi^{-}$ is -0.141$\pm$0.040$\pm$0.018 from LHC experiments \cite{PL2014}. Our result is -0.18$\pm$0.04$\pm$0.003 for the $B^{-}\rightarrow \phi \pi^{-}\rightarrow K^{+}K^{-}\pi^{-}$ decay without the mixing from the our framework under the region of $m_{K^{+}K^{-}}$$<$1.5 $\textrm{GeV}^2$, which is consistent with the experimental results.

Except above error in the B part of IV, we also need to consider the the narrow width approximation (NWA), where exists a factorization relation. In the former calculation, we take the method of the quasi-two-body decay process and also consider the resonance effect where the vector meson can further decay into two hadrons \cite{4Lu2019}. Hence, the total amplitude form can be regarded as two parts after we taking the quasi-two-body decay process and the degree of approximation can be connected by the term of $\eta_{R}$ \cite{plb2021}.
The formalisms of decay amplitudes contain the Breit-Wigner shapes which depend on the parameters of invariant mass $m_{K^{+}K^{-}}$ associated with the Dalitz plot. We integrate over invariant mass $m_{K^{+}K^{-}}$ in order to obtain localized value of CP asymmetry by quasi-two-body approximation. The introduced CP asymmetry can provide a favorable theoretical support for the experimental exploration in the future.

\section{Acknowledgments}
This work was supported by  Natural Science Foundation of Henan (Project No.232300420115) and National Natural Science Foundation of China (Project No.12275024).

%\newpage

\end{spacing}
\end{document}